\documentclass[aps,prl,twocolumn,showpacs]{revtex4}

\usepackage{graphicx}
\usepackage{dcolumn}
\usepackage{bm}

\begin{document}


\title{
Nonequilibrium chaos of disordered nonlinear waves
}

\author{Ch.~Skokos} \email{hskokos@auth.gr} \affiliation{Physics
  Department, Aristotle University of Thessaloniki, GR-54124,
  Thessaloniki, Greece}
\author{I.~Gkolias} \affiliation{Physics Department, Aristotle
   University of Thessaloniki, GR-54124, Thessaloniki, Greece}
\author{S.~Flach} \affiliation{New Zealand Institute for Advanced
  Study, Centre for Theoretical Chemistry and Physics, Massey
  University, Auckland, New Zealand}

\date{\today}

\begin{abstract}
  Do nonlinear waves destroy Anderson localization? Computational and
  experimental studies yield subdiffusive nonequilibrium wave packet
  spreading. Chaotic dynamics and phase decoherence assumptions are
  used for explaining the data. We perform a quantitative analysis of
  the nonequilibrium chaos assumption, and compute the time dependence
  of main chaos indicators - Lyapunov exponents and deviation vector
  distributions. We find a slowing down of chaotic dynamics, which
  does not cross over into regular dynamics up to the largest observed
  time scales, still being fast enough to allow for a thermalization
  of the spreading wave packet.  Strongly localized chaotic spots
  meander through the system as time evolves.  Our findings confirm
  for the first time that nonequilibrium chaos and phase decoherence
  persist, fueling the prediction of a complete delocalization.
\end{abstract}

\maketitle


\paragraph{Introduction.}
In one and two dimensions, wave packets of noninteracting particles
subject to a random potential on a lattice do not propagate due to
exponential Anderson localization of the corresponding eigenstates
\cite{anderson1958, kramer1993}. Anderson localization is a highly
intriguing wave phenomenon, and has been recently probed in
experiments on ultracold atomic gases in optical potentials
\cite{billy2008,roati2008}.  The wave localization effect is
completely relying on keeping the phase coherence of participating
waves.  The presence of interaction between the particles may change
this picture qualitatively. For many weakly interacting particles this
is often taken into account by adding nonlinear terms to the linear
wave equations of the noninteracting particles \cite{pitaevsky2003}.

Numerical simulations of wave packets propagating in random lattice
potentials showed the destruction of localization and a subdiffusive
growth of the second moment of the wave packet in time as $t^\gamma$
\cite{kopidakis2008,pikovsky2008,garcia2009,flach2009,skokos2009,veksler2009,
  skokos2010,laptyeva2010,bodyfelt2011,laptyeva2012,gligoric2013}. In
particular, it was predicted that, for asymptotically large $t$, the
coefficient $\gamma$ should converge to $1/(1+\sigma d)$ in the
so-called `weak chaos' regime, where $d$ is the dimension of the
lattice, and $\sigma+1$ is the exponent of the nonlinear term in the
wave equation (note that usual two-body interactions yield cubic
nonlinear terms with $\sigma=2$)
\cite{flach2009,skokos2009,flach2010}.  A transient `strong chaos'
regime was also predicted \cite{flach2010} and observed
\cite{laptyeva2010,bodyfelt2011} where $\gamma=2/(2+\sigma
d)$. Remarkably the onset of subdiffusive spreading was also
experimentally detected for interacting ultracold atoms in optical
potentials \cite{lucioni2011}.

The main dynamical origin of the observed subdiffusion is believed to
be deterministic chaos. Indeed, assume that a wave packet which is
exciting a few lattice sites (or more) is not spreading.  Then the
dynamics of this trapped excitation can be described within a
Hamiltonian system with a finite number of degrees of freedom, which
is nonlinear, and typically not integrable \cite{gutzwiller1990}.
Excluding very weak nonlinearities and the Kolmogorov-Arnold-Moser
(KAM) regime, the dynamics should be chaotic. Since deterministic
chaos deteriorates correlations and leads to a decoherence of the wave
phases, the main ingredient of Anderson localization is lost, and the
wave packet can not be anymore localized \cite{rayanov2013}. Therefore
one contradicts the initial assumption of persistence of
localization. The wave packet can not keep its localization, but will
spread.  From this it also follows, that for weak enough nonlinearity
the initially excited wave packet is in a KAM regime, where there is a
finite probability to start on a chaotic trajectory, but with a finite
complementary probability to end up on a regular trajectory which
enjoys phase coherence and localization \cite{gutzwiller1990}.  Indeed
it was recently shown that the probability for chaos is practically
equal to one above a nonlinearity threshold, and less than one below
that threshold, tending towards zero in the limit of vanishing
nonlinearity, thus restoring Anderson localization in this
probabilistic sense \cite{ivanchenko2011}.  The microscopic origin of
chaos is expected to be hidden in the chaotic seeds of nonlinear
resonances, which will fluctuate in time and space as the wave packet
spreads \cite{krimer2010}.

To summarize, two main assumptions - the deterministic chaoticity of
the wave packet dynamics, and the space-time fluctuation of the
chaotic seeds - have to be confirmed to provide solid grounds for
subdiffusion theories.  A study of the interrelation between the
appearance of chaos for nonlinear waves in the disordered
Schr\"odinger equation was performed by Mulansky et
al.~\cite{mulansky2009}.  Various relations between mode energies and
capabilities to reach thermal equilibrium were studied for small
systems, but not for the nonequilibrium wave packet spreading
situation.  In \cite{TSL13} it has been shown that chaoticity does not
necessarily imply thermalization for small-size disordered lattices.
Michaely and Fishman \cite{michaely2012} recently studied the temporal
and frequency characteristics of effective nonlinear forces inside the
wave packet, and concluded that these forces show sufficient
randomness to qualify as effective noise terms - which is thought to
be a result of deterministic chaos (but not a direct proof of its
existence). Similarly Vermersch and Garreau \cite{vermersch2012}
measured a spectral entropy, which however is only a very rough
indicator for chaos. They also attempted to measure Lyapunov
exponents, but only on short times. Moreover, all these attempts do
not account for the {\sl temporal dependence} of chaos
strength. Indeed, the more the wave packet spreads, the weaker the
chaos should become, since densities decrease. This is also reflected
in the fact that the packets spread subdiffusively. Therefore we need
a temporal resolution of the chaos indicators. That is what we will
present in this work.

\paragraph{Model, equations and methods of analysis.}
The spreading of wave packets was numerically studied in a number of
classes of wave equations. These equations share a surprising
universality in that only the dimensionality of the lattice and the
nonlinearity power $\sigma$ influence the value of the exponent
$\gamma$.  Therefore we choose a chain of coupled anharmonic
oscillators with random harmonic frequencies which belongs to the
class of quartic Klein-Gordon (KG) lattices.  This model is
dynamically very similar to nonlinear Schr\"odinger equations with
random potentials for small densities
\cite{flach2009,skokos2009,laptyeva2010,bodyfelt2011,laptyeva2012,ivanchenko2011,laptyeva2012-2}.
The Hamiltonian of the quartic KG chain of coupled anharmonic
oscillators with coordinates $u_l$ and momenta $p_l$ is
\begin{equation}
  \mathcal{H}_{K}= \sum_l  \frac{p_l^2}{2} +\frac{\tilde{\epsilon}_l}{2} u_l^2 +
  \frac{1}{4} u_l^4+\frac{1}{2W}(u_{l+1}-u_l)^2\;.
\label{RQKG}
\end{equation}
The equations of motion are $\ddot{u}_l = - \partial \mathcal{H}_{K}
/\partial u_l$, and $\tilde{\epsilon}_l$ are chosen uniformly from the
interval $\left[\frac{1}{2},\frac{3}{2}\right]$. The value of
$\mathcal{H}_{K}$ serves as a control parameter of the system's
nonlinearity. In the absence of the quartic term in (\ref{RQKG}) the
ansatz $u_l(t) = A_l {\rm e}^{i\omega t}$ yields the linear eigenvalue
problem $\lambda A_l = \epsilon_l A_l -(A_{l+1} + A_{l-1})$ with
$\epsilon_l=W(\tilde{\epsilon}_l -1)$ and $\lambda = W(\omega^2 - 1) -
2$.  This eigenvalue problem corresponds precisely to the well known
Anderson localization in a one-dimensional chain with diagonal
disorder \cite{anderson1958,kramer1993}.

We analyze normalized energy distributions $\epsilon_{l} \geq 0$ using
the second moment $m_2= \sum_{l} (l-\bar{l})^2 \epsilon_{l}$, which
quantifies the wave packet's degree of spreading and the participation
number $P=1 / \sum_{l} \epsilon_{l}^2$, which measures the number of
the strongest excited sites in $\epsilon_{l}$.  Here $\bar{{l}} =
\sum_{l} l \epsilon_{l}$ and $\epsilon_{l}\equiv
h_{l}/\mathcal{H}_{K}$ with $h_{l} = p_l^2/2 +\tilde{\epsilon}_l u_l^2
/2 + u_l^4/4+(u_{l+1}-u_l)^2/4W$. During the wave packet evolution we
further estimate the maximum Lyapunov exponent (mLE) $\Lambda_1$ as
the limit for $t \rightarrow \infty$ of the quantity
$\Lambda(t)=t^{-1} \ln ( \| \vec{v}(t)\| / \| \vec{v}(0)\|)$, often
called {\sl finite time mLE} \cite{BGGS80a,BGGS80b,S10}. $\vec{v}(0)$,
$\vec{v}(t)$ are deviation vectors from a given trajectory, at times
$t=0$ and $t>0$ respectively, and $\| \cdot \|$ denotes the usual
vector norm. $\Lambda(t)$ is a widely used chaos indicator. It tends
to zero in the case of regular motion as $\Lambda^r(t)\sim t^{-1}$
\cite{BGS76,S10}, while it tends to nonzero values for chaotic
motion. Its inverse, $T_L$, is the characteristic timescale of the
studied dynamical system, the so-called Lyapunov time. It quantifies
the time needed for the system to become chaotic.  The vector
$\vec{v}(t)$ has as coordinates small deviations from the studied
trajectory in positions and momenta ($v_i=\delta u_i$, $v_{i+N}=\delta
p_i$, $1\leq i \leq N$, $N$ being the total number of lattice
sites). Its time evolution is governed by the so-called variational
equations. In our study we also compute normalized deviation vector
distributions (DVDs) $w_l= (v_l^2+v_{l+N}^2)/\sum_l(v_l^2+v_{l+N}^2)$.

We use the symplectic integrator SABA$_2$ with corrector
\cite{LR01,skokos2009} for the integration of the equations of motion,
and its extension according to the so-called tangent map method
\cite{SG10,GS11,GES12} for the integration of the variational
equations. We considered lattices with $N=1000$ to $N=2000$ sites in
our computations, in order to exclude finite-size effects in the
evolution of the wave packets, and an integration time step
$\tau=0.2$, which kept the relative energy error always less than
$10^{-4}$.

\begin{figure*}
\includegraphics[width=\linewidth]{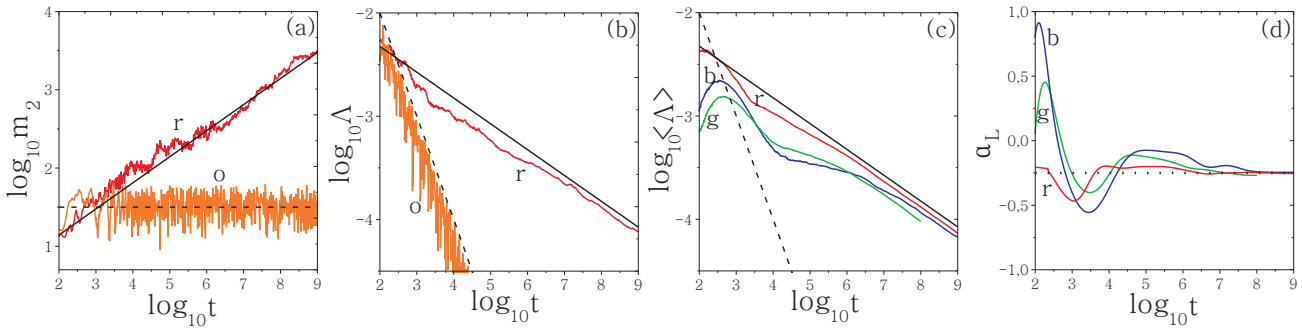}
\caption{(Color online) (a) Time evolution of the second moment $m_2$
  for one disorder realization of an initially single site excitation
  with $E=0.4$, $W=4$ (Case I), in $\log - \log$ scale (red (r)
  curve). The orange (o) curve corresponds to the solution of the
  linear equations of motion, where the term $u_l^4$ in
  Eq.(\ref{RQKG}) is absent. Straight lines guide the eye for slopes
  $1/3$ (solid line) and $0$ (dashed line).  (b) Time evolution of the
  finite time maximum Lyapunov exponent $\Lambda$ (multiplied by 10
  for the orange (o) curve) for the trajectories of panel (a) in $\log
  - \log$ scale. The straight lines guide the eye for slope $-1$
  (dashed line), and $-1/4$ (solid line). (c) Time evolution of the
  averaged $\Lambda$ over 50 disorder realizations for the `weak
  chaos' cases I, II and III [(r) red; (b) blue; (g) green] (see text
  for more details). Straight lines guide the eye for slopes $-1$ and
  $-1/4$ as in panel (b). (d) Numerically computed slopes $\alpha_L$
  of the three curves of panel (c). The horizontal dotted line denotes
  the value $-1/4$.}.
\label{fig1}
\end{figure*}

\paragraph{Results.}
In Fig.\ref{fig1}(a) we first study the single trajectory of a single
site excitation with total energy $E=0.4$ and $W=4$ (case I) which is
known to evolve in the asymptotic regime of `weak chaos'
\cite{flach2009,skokos2009,laptyeva2010}. We show the time dependence
of the second moment (red curve) and observe the expected subdiffusive
growth $m_2 \sim t^{1/3}$. The simulation of a single site excitation
in the absence of nonlinear terms (orange curve) corresponds to
regular motion and Anderson localization is observed. In
Fig.\ref{fig1}(b) we plot the time dependence of $\Lambda(t)$ for the
two cases of Fig.\ref{fig1}(a). At variance to the $t^{-1}$ decay for
the regular nonchaotic trajectory (orange curve), the observed decay
for the `weak chaos' orbit is much weaker and well fitted with
$\Lambda \sim t^{-1/4}$ (red curve).

We substantiate the findings of Fig.\ref{fig1}(b) by averaging
$\log_{10} \Lambda$ over 50 realizations of disorder and extending to
two more `weak chaos' parameter cases with initial energy density
$\epsilon=0.01$ distributed evenly among a block of 21 central sites
for $W=4$ (case II) and 37 central sites for $W=3$ (case III). All
cases show convergence towards $\Lambda \sim t^{-1/4}$
(Fig.\ref{fig1}(c)). We further differentiate the curves in
Fig.\ref{fig1}(c) following the approach used in
\cite{laptyeva2010,bodyfelt2011}, estimate their slope
$\alpha_L=\frac{d(\log_{10}\Lambda(t))}{d\log_{10}t}$, and show the
result in Fig.\ref{fig1}(d) which underpin the above findings.
Therefore
\begin{equation}
  \Lambda(t)  \sim t^{-1/4} \gg \frac{1}{t} \;.\label{chaosreg}
\end{equation}

So far we have clear numerical proof that the dynamics inside the
spreading wave packet remains chaotic up to the largest simulation
times, without any tendency towards regular dynamics. As expected, the
chaoticity diminishes in time due to the decrease of the energy
density inside the spreading wave packet.  Therefore, complete
chaotization and randomization of phases inside the wave packet will
take more time, the more the packet spreads.  To substantiate the
assumptions needed for subdiffusive spreading theories, we will
compare the Lyapunov time $T_L$ (estimated by $1/\Lambda$) with the
time scales which characterize the subdiffusive spreading. A first
time scale of this kind, $T_D$, can be obtained from the growth of the
second moment $m_2 \sim t^{1/3}$.  It follows that the inverse of this
timescale, i.e.~the effective diffusion coefficient $D$, is a function
of the densities, and decays in time as \cite{laptyeva2012-2}
\begin{equation}
  \frac{1}{T_D}=D \sim t^{-2/3}\;,\;\Lambda \gg D \;,\; \frac{T_D}{T_L}\sim t^{5/12}\;.
  \label{lambda-D}
\end{equation}
A second time scale can be obtained by estimating a spreading time
$T_s$ given by the increase of $P$ by one (site), i.e.~$T_s \sim
1/\dot{P}$ (with $P\sim t^{1/6}$
\cite{flach2009,skokos2009,bodyfelt2011}). Therefore
\begin{equation}
  \frac{1}{T_s} \sim t^{-5/6}\;,\;\Lambda \gg \frac{1}{T_s}\;,\; \frac{T_s}{T_L} \sim t^{7/12}\;.
  \label{lambda-Ts}
\end{equation}
As it follows from
Eqs.(\ref{chaosreg},\ref{lambda-D},\ref{lambda-Ts}), the dynamics
remains chaotic, and the chaoticity time scale is always shorter than
the spreading time scales, and their ratio diverges as a power law. We
thus confirm for the first time the assumption about persistent and
fast enough chaoticity needed for subdiffusive spreading theories.

A second very important assumption for subdiffusive spreading theories
is based on the fact that chaoticity is induced by nonlinear
resonances inside the wave packet, which are the seeds of
deterministic chaos and {\sl have to meander} through the packet in
the course of evolution \cite{skokos2009,krimer2010}. Indeed, assume
that their spatial position is fixed. Then such seeds will act as
spatially pinned random force sources on their surrounding. The noise
intensity of these centers will decay in time. At any given time the
exterior of the wave packet is then assumed to be approximated by the
linear wave equation part which enjoys Anderson localization. However,
even for constant intensity it was shown \cite{aubry2009} that the
noise will not propagate into the system due to the dense discrete
spectrum of the linear wave equation. Therefore the wave packet can
only spread if the nonlinear resonance locations meander in space and
time.

We visualize the motion of these chaotic seeds by following the
spatial evolution of the deviation vector used for the computation of
the mLE. This vector tends to align with the most unstable direction
in the system's phase space. Thus, monitoring how its components on
the lattice sites evolve we can identify the most chaotic spots. Large
DVD values tell us at which sites the sensitivity on initial
conditions (which is a basic ingredient of chaos) is larger.

\begin{figure}
\includegraphics[scale=.55]{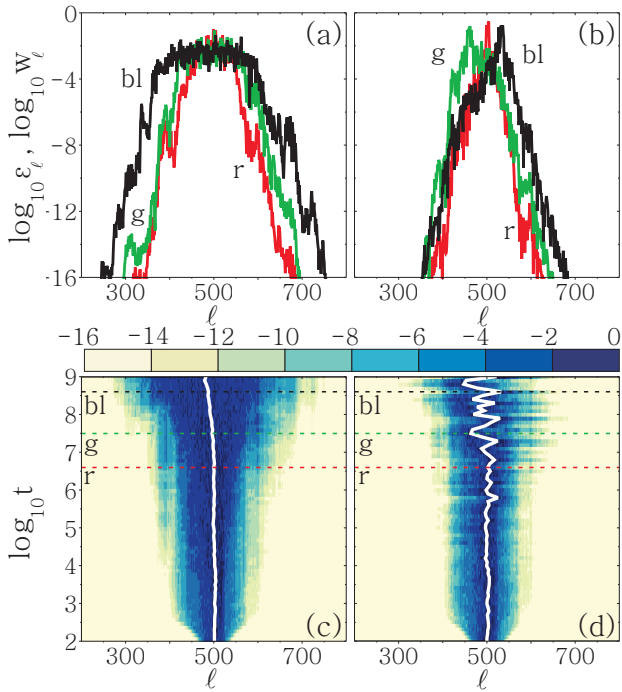}
\caption{(Color online) The dynamics of an individual trajectory of
  case I. Normalized (a) energy ($\epsilon_l$) and (b) deviation
  vector ($w_l$) distributions at $t=4\times 10^{6}$, $t=3\times
  10^{7}$, $t=4 \times 10^8$ [(r) red; (g) green; (bl) black].  Time
  evolution of (c) the energy distribution and (d) the DVD for the
  realization of panel (a) in $\log_{10}$ scale.  The position of the
  distribution's mean position is traced by a thick white curve. The
  times at which the distributions of panels (a) and (b) are taken are
  denoted by straight horizontal lines in (c) and (d).}
\label{fig2}
\end{figure}

In Fig.\ref{fig2}(a) we plot the energy density distribution for an
individual trajectory of case I at three different times
$t\approx10^6,\,10^7,\,10^8$ and in Fig.\ref{fig2}(b) the
corresponding DVD.  We obtain that the energy densities spread more
evenly over the lattice the more the wave packet grows. At the same
time the DVD stays localized, but the peak positions clearly meander
in time, covering distances of the order of the wave packet width. The
full time evolution of the energy density and the DVD is shown in
Figs.\ref{fig2}(c,d) together with the track of the distribution's
mean position (central white curve). While the energy density
distribution shows a modest time dependence of the position of its
mean, the DVD mean position is observed to perform fluctuations whose
amplitude increases with time.

\paragraph{Summary and discussion.}

We computed nonequilibrium chaos indicators of the spreading of wave
packets in disordered lattices.  For the first time we find that chaos
not only exists, but also persists. Using a set of observables we are
able to show that the slowing down of chaos does not cross over into
regular dynamics, and is at all times fast enough to allow for a
thermalization of the wave packet. Moreover the monitoring of the
spatio-temporal dynamics of hot chaotic spots yields an increase in
their spatial fluctuations - in accord with previous unproven
assumptions.

The mLE decreases when the wave packet spreads, since the energy
density decreases as well. Nevertheless, the mLE follows a completely
different power law as compared to the case of regular motion. This
signals the lack of sticking to regular structures in phase space, as
conjectured recently \cite{JKA10,A11}.

We also studied the spatial evolution of the deviation vector
associated with the mLE. The corresponding distributions remain
localized with a very pointy profile. This observation supports
theoretical assumptions that in the `weak chaos' regime only few
nonlinear resonances appear at a time.  The mean position of these
distributions performs random oscillations, whose amplitude increases
as the wave packet spreads. These oscillations result in a homogeneity
of chaos inside the packet, i.e.~in thermalization.

All these findings clearly show that nonlinear wave packets spread in
random potentials due to deterministic chaos and dephasing.  Moreover,
wave packets first thermalize, and only later perform subdiffusive
spreading.  That is a basic prerequisite for the existing theoretical
description of energy spreading in disordered nonlinear lattices, and
the applicability of nonlinear diffusion equations
\cite{laptyeva2012-2,mulansky2009b,mulansky2013,lucioni2013}.

\paragraph{Acknowledgments.}
Ch.S. and S.F. thank the Max Planck Institute for the Physics of
Complex Systems in Dresden, Germany for its hospitality during their
visits, when part of this work was carried out. Ch.S.~was supported by
the Research Committee of the Aristotle University of Thessaloniki
(Prog.~No 89317), and by the European Union (European Social Fund -
ESF) and Greek national funds through the Operational Program
``Education and Lifelong Learning'' of the National Strategic
Reference Framework (NSRF) - Research Funding Program:
``THALES. Investing in knowledge society through the European Social
Fund''. The HellasGrid infrastructure was used for the numerical
simulations.


\end{document}